\documentclass[%
 reprint,
 amsmath,amssymb,
 aps,
prl]{revtex4-2}

\usepackage{graphicx}% Include figure files
\usepackage{dcolumn}% Align table columns on decimal point
\usepackage{bm}
\usepackage{bbold}
\usepackage{braket}
\usepackage{xcolor}
\usepackage[final]{changes}
\begin{document}

\preprint{APS/123-QED}

\title{Optical Indistinguishability via Twinning Fields}% Force line breaks with \\

\author{Gerard McCaul}
\email{gmccaul@tulane.edu}
\affiliation{Tulane University, New Orleans, LA 70118, USA}

\author{Alexander F. King}
\affiliation{Tulane University, New Orleans, LA 70118, USA}

\author{Denys I. Bondar}
\affiliation{Tulane University, New Orleans, LA 70118, USA}

\date{\today}% It is always \today, today,

\begin{abstract}
 Here we introduce the concept of the twinning field -- a driving electromagnetic pulse that induces an identical optical response from two distinct materials.  We show that for \added{a large class of} pairs of generic many-body systems, a twinning field which renders the systems \emph{optically indistinguishable} exists. \added{The conditions under which this field exists are derived, and this analysis is supplemented by numerical calculations of twinning fields for both the 1D Fermi-Hubbard model, and tight-binding models of graphene and hexagonal Boron Nitride}. The existence of twinning fields may lead to new research directions in non-linear optics, materials science, and quantum technologies.
\end{abstract}

\maketitle

%\tableofcontents

\paragraph{Introduction:-}  As our understanding of the physical world has progressed to mastery over it, it has become apparent that the qualities which define a material at equilibrium may be modified under driving. This phenomenon underpins both quantum simulation \cite{RevModPhys.86.153} and Floquet engineering \cite{Floquet,Floquet2,Floquet3}. One of the principal goals of quantum control theory \cite{Werschnik2007} is the specification of the driving fields necessary either to steer a system to some desired state \cite{Serban2005,PhysRevLett.106.190501,PhysRevA.37.4950, RevModPhys.80.117, Glaser2015,PRXQuantum.2.010101}, or fulfill a pre-specified condition on its expectations \citep{Kosloff1992,Koslofflocalcontrol}. 

Using tracking quantum control  \cite{Ong1984,Clark1985,PhysRevA.72.023416,PhysRevA.98.043429,PhysRevA.84.022326,Campos2017, doi:10.1063/1.1582847,doi:10.1063/1.477857}, recent work has demonstrated that almost arbitrary control over the optical response of a large class of solid-state systems can be achieved \cite{tracking1,tracking2}. One consequence of this is that two specially tailored driving fields will induce an identical response from two distinct systems. Given the essential malleability of quantum systems under driving, one might ask whether it is possible to fulfil the stronger condition of obtaining identical responses using the \emph{same} driving field on each system. Put differently, do there exist fields for which a pair of systems' response are indistinguishable? 

 Consider two distinct systems $\ket{\psi_1}$ and $\ket{\psi_2}$, with an identical control field impinging on each of them (see Fig. \ref{fig:twinning_field}). Each system will generate an optical response $J^{(k)}(t)$, and if $J^{(1)}(t)=J^{(2)}(t)$ for all times $t$, then the driving field is what we term a `twinning field', and the systems are \emph{optically indistinguishable}.  

 In the regime of linear response, this may initially appear trivial, as many systems possess extremely similar absorption and emission spectra over a broad range of frequencies (e.g., large organic molecules \cite{lichtman2005fluorescence}). Indeed, such is the closeness of these systems' response that quantum control \cite{brixner2001photoselective,photonicreagents,oddrabitz,li2005optimal, rondi2012coherent, roslund2011resolution, PhysRevLett.102.253001}  (including tracking control \cite{2010.13859}) must be exploited to accurately detect these systems. Of course, similar is not \emph{identical}, \added{and purely linear response would require identical susceptibilities for identical responses}. In general however, materials also have a non-linear component to differentiate them (see, e.g., \cite{drobizhev2011two}). In fact many important phenomena -- e.g. high harmonic generation \citep{Ghimire2012, Ghimire2011a, Murakami2018}, the workhorse of attosecond physics \cite{Corkum2007,RevModPhys.81.163,Li2020} -- explicitly rely on optical non-linearities  \cite{boyd_nonlinear_2008}.

 \begin{figure}
\begin{center}
\includegraphics[width=1\columnwidth]{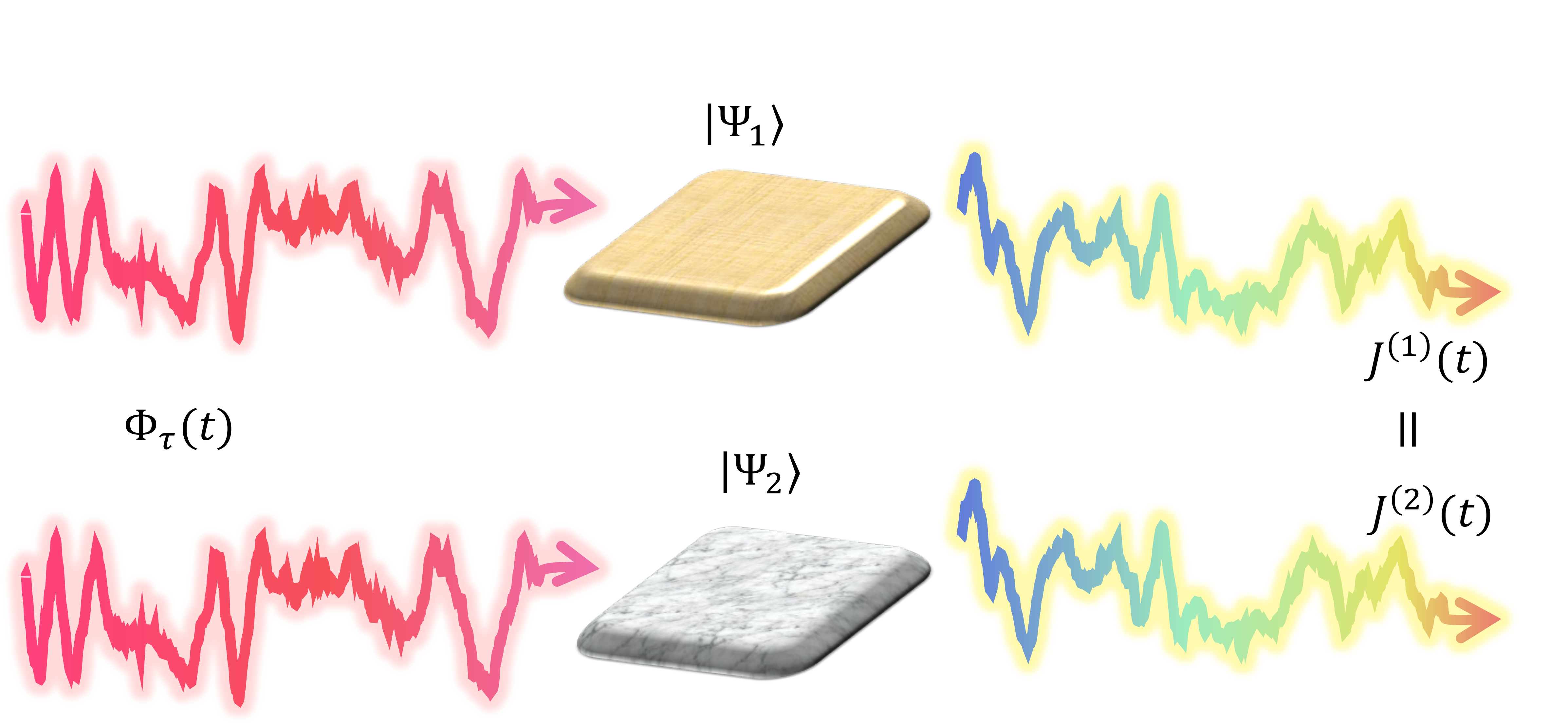}\end{center}\caption{Ordinarily, distinct systems will have different responses to the same driving field. A twinning field relates a pair of systems as the field under which the optical response $J^{(k)}(t)$ of each system is identical.}
\label{fig:twinning_field}
\end{figure}
 
 For this reason, true optical indistinguishability must be considered in the context of the non-linear response that arises in a fully quantum treatment of materials. Some preliminary hints of this indistinguishability have been observed experimentally, specifically in the non-uniqueness of the parametrisation of second-order nonlinear spectra \cite{Busson2009}.
 To date however, a theoretical justification for such results has been lacking. In this manuscript we address this issue and present a framework for achieving driven indistinguishability. The main result is a demonstration that for any pair of generic many-electron systems on a lattice, there always exists a twinning field which will elicit an identical response from each system. Furthermore, the conditions under which this field is unique are established. This framework is then extended by deriving a general twinning field which renders an arbitrary expectation identical between systems. Finally we discuss the physical implications of twinning fields, and their potential utility.     

\paragraph{Results:- \label{sec:Model}} \added{Here we outline the derivation of the twinning field for 1D systems, while the more general case is discussed in the supplementary material \cite{supplement} which includes Refs. \cite{kentnagle2011,geraldfolland2007, Rabitzmultiple}. Here we consider two many-electron systems on a lattice, labeled by $k=1,2$. Each has a potential $\hat{U}^{(k)}$ due to electron-electron interactions.} Both systems are excited by an identical laser pulse, described under the dipole approximation  by the Peierls phase $\Phi(t)$ \cite{Peierls1933,PhysRevA.95.023601}. Such systems' evolution will then be determined by the Hamiltonian (in atomic units) \cite{tracking2,fabianessler2005}:
\begin{align}
\hat{H}^{(k)}(t)=  -t^{(k)}_{0}\sum_{j,\sigma}\text{\ensuremath{\left({\rm e}^{-i\Phi\left(t\right)}\hat{c}_{j\sigma}^{\dagger}\hat{c}_{j+1\sigma}+ \rm{h.c.}\right)}} +\hat{U}^{(k)}, \label{eq:Hamdiscrete} 
\end{align}
where $\hat{c}_{j\sigma}$ is the fermionic annhilation operator (acting on the appropriate system) for site $j$ and spin $\sigma$, satisfying the anticommutation relation $\{\hat{c}^\dagger_{j\sigma},\hat{c}_{j^{\prime}\sigma^\prime}\}=\delta_{\sigma \sigma^\prime}\delta_{jj^{\prime}}$, while $t^{(k)}_0$ is the hopping parameter describing the kinetic energy of the electrons.

Consider a typical example of an optically driven current. The current operator $\hat{J}^{(k)}$ is defined from a continuity equation for the electron density \cite{tracking1,tracking2}. Provided all number operators $\hat{n}_{j\sigma}=\hat{c}^\dagger_{j\sigma}\hat{c}_{j\sigma}$ commute with $\hat{U}^{(k)}$, each system's current operator has the form \cite{Silva2018}:  
\begin{equation}
    \hat{J}^{(k)}(t)=-iat^{(k)}_{0}\sum_{j,\sigma}\left({\rm e}^{-i\Phi\left(t\right)}\hat{c}_{j\sigma}^{\dagger}\hat{c}_{j+1\sigma}-{\rm h.c.}\right),\label{eq:currentoperator}
\end{equation}
where $a^{(k)}$ is the lattice constant. It is important to note that the current expectation $J^{(k)}(t)=\braket{\psi_k(t) |\hat{J}^{(k)}(t)|\psi_k(t)}$ depends only implicitly on $\hat{U}^{(k)}$ through the evolution of $\ket{\psi_k(t)}$, significantly simplifying expressions.  
  
Having dispensed with this preamble, we come to our main topic of investigation. For two systems $\ket{\psi_1}$ and $\ket{\psi_2}$ with potentials $\hat{U}^{(1)}$ and $\hat{U}^{(2)}$, does there exist a twinning field $\Phi_\tau(t)$ such that $J^{(1)}(t)=J^{(2)}(t)$, making the response of one system indistinguishable from the other? 

To establish the existence of this field, we first express the nearest-neighbour expectation of each system in a polar form
\begin{align}
\hat{K}&= \sum_{j,\sigma}\hat{c}_{j\sigma}^{\dagger}\hat{c}_{j+1\sigma}, &\label{eq:K} \\
K(\psi_k)&=\left\langle \psi_k (t) \left|\hat{K}\right| \psi_k (t)  \right\rangle =R\left(\psi_k\right){\rm e}^{i\theta\left(\psi_k\right)}. \label{neighbourexpectation}
\end{align}
Note that in both this and later expressions, the argument $\psi_k$ indicates that the expression is a functional of  $\ket{\psi_k}\equiv \ket{\psi_k(t)}$. \added{We emphasise that this functional will have a well defined value for any state that has been obtained through evolution under the Hamiltonian given in Eq.~\eqref{eq:Hamdiscrete}.} Using this, we may express the response expectation $J^{(k)}$ directly as
\begin{align}
J^{(k)}\left(t\right)= & -i a t_{0} R\left(\psi_k\right)\left({\rm e}^{-i\left[\Phi\left(t\right)-\theta\left(\psi_k\right)\right]}-{\rm e}^{i\left[\Phi\left(t\right)-\theta\left(\psi_k\right)\right]}\right)\nonumber \\
= & -2 a t_{0} R \left(\psi_k\right)\sin(\Phi\left(t\right)-\theta\left(\psi_k\right)).\label{eq:currentexpectation}
\end{align}
It is straightforward to equate the currents $J^{(1)}(t)$ and $J^{(2)}(t)$ and obtain an expression for the twinning field $\Phi_\tau$ in terms of the expectations of the two systems:
\begin{align}
    \Phi_\tau(t) &=\arctan\left(\xi(\psi_1,\psi_2)\right) \label{eq:twinningfield}\\ 
    \xi(\psi_1,\psi_2)&=\frac{\lambda R(\psi_1)\sin(\theta(\psi_1))-R(\psi_2)\sin(\theta(\psi_2))}{ \lambda R(\psi_1)\cos(\theta(\psi_1))-R(\psi_2)\cos(\theta(\psi_2))},
\end{align}
where $\lambda=\frac{a^{(1)}t^{(1)}_0}{a^{(2)}t^{(2)}_0}$. \added{Critically, in this 1D case one is able to obtain a closed form for $\Phi_\tau (t)$, such that the existence of this field can be assessed purely by considering its right hand side.} 
Given both the range and domain of $\arctan$ extends over the reals and $\xi(\psi_1,\psi_2)$ is real by definition \added{(and has a definite value for any pair of states)}, we can immediately conclude that \emph{a twinning field between any two systems described by Eq.~\eqref{eq:Hamdiscrete} always exists.} 

\added{An important caveat to this statement is that the predicted twinning field may be identically zero depending on the initial states of the twinned systems. For example, if we attempt to twin two systems of non-interacting electrons ($\hat{U}^{(k)}=0$), then $\hat{K}$ commutes with the Hamiltonian and $K(\psi_k)$ is constant. If the systems are evolved from their ground state, $\theta(\psi_k)=0$, and by Eq.~\eqref{eq:twinningfield}, $\Phi_\tau (t)=0$. This scenario is consistent with the impossibility of twinning fields in linear optics, and can be avoided by having at least one of the system pair have a non-zero potential and hence a nonlinear response. Furthermore, while an equation for $\Phi_\tau$ can still be obtained in higher dimensions, in general it will not be of a closed form, and therefore a twinning field is not guaranteed to exist. The additional requirements for a twinning field to exist in this scenario are detailed in the supplementary material  \cite{supplement}.}

\paragraph{Illustrations:-} 
Here we provide examples of twinning fields for systems described by the Fermi-Hubbard model of strongly interacting electrons. In this case, each system has an onsite potential described by \cite{fabianessler2005,Silva2018}: 
\begin{equation}
\label{eq:onsitepotential}
    \hat{U}^{(k)} =U^{(k)}\sum_j \hat{c}_{j\uparrow}^{\dagger}\hat{c}_{j\uparrow}\hat{c}_{j\downarrow}^{\dagger}\hat{c}_{j\downarrow} 
\end{equation}
where $U^{(k)}$ parametrises the energy of the electron-electron repulsion. Systems'  equilibrium properties are determined by the ratio $U^{(k)}/t^{(k)}_0$. Despite the simplicity of the potential, this model is rich in nontrivial behaviour, including topological \cite{Le2020,PhysRevB.102.174314} and superconducting phases in 2D \cite{PhysRevLett.95.237001,PhysRevLett.110.216405}. The Fermi-Hubbard model is  computationally challenging and a complete understanding of its dynamics  is  believed to require a quantum computer \cite{mcardle_quantum_2020}. It also exhibits a highly non-linear optical response \cite{Keimer2017, Silva2018,Tokura2018}, and therefore provides a suitable platform for numerical calculations of twinning fields. 

Here we consider an $L=10$ site chain with periodic boundary conditions and an average of one electron per site. For the sake of simplicity, in both systems we use the lattice constants $a^{(1)}=a^{(2)}=4$~\AA, with a hopping parameter of $t_0\equiv t^{(1)}_0=t_0^{(2)}=0.52$~eV. To avoid the trivial solution of $\Phi_\tau(t)=0$, each system (initially in the ground state) is first pumped by a single cycle of a transform-limited field. Specifically, an enveloped sine-wave is used  with an amplitude of $E_0=10$~MV/cm and frequency $\omega_0=32.9$~THz. All calculations were performed using exact diagonalization via the QuSpin Python package \cite{10.21468/SciPostPhys.7.2.020,SciPostPhys.2.1.003}. 
 \begin{figure}
\begin{center}
\includegraphics[width=1\columnwidth]{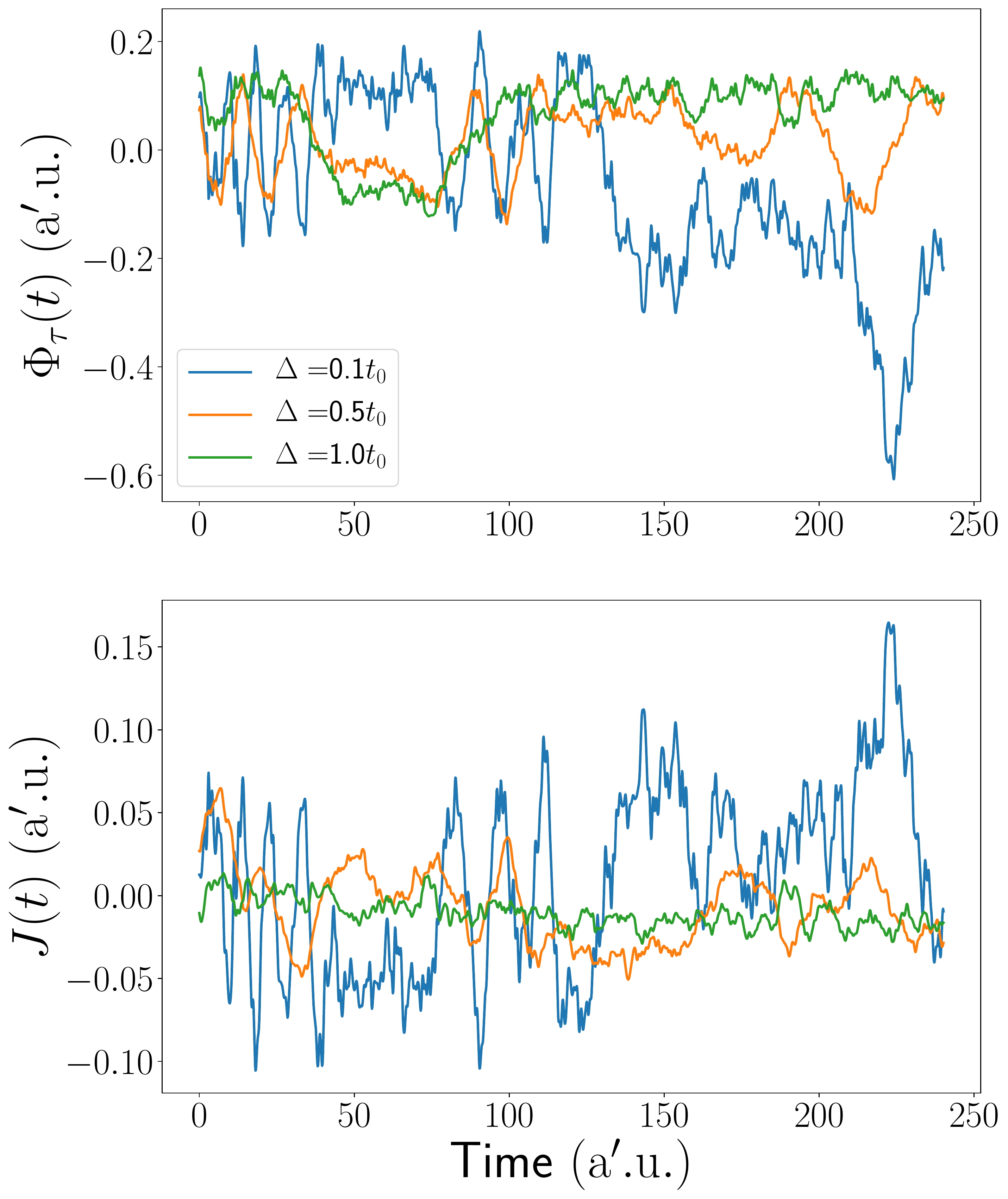}\end{center}\caption{Top panel: twinning fields and the accompanying optical response for three pairs of systems, following an initial pump pulse. Bottom panel: The current resulting from the application of the twinning field, which is identical in each pair of systems. In all cases $U^{(1)}=0.5 t_0$, while $U^{(2)}=U^{(1)}+\Delta$, where $\Delta$ is varied for each of the three pairs. ${\rm a^\prime.u.}$ are atomic units with energy normalised to $t_0$. }
\label{fig:lowU}
\end{figure}

\added{Figure~\ref{fig:lowU} shows examples of calculated twinning fields and the accompanying responses they generate for several pairs of systems. These pairs are parametrised by $\Delta$, using $U^{(1)}=0.5 t_0$, while $U^{(2)}=U^{(1)}+\Delta$. Applying the twinning field calculated at each time, we find the current in each pair of systems is identical, as expected.}

% \deleted{As Fig.~\ref{fig:highU}- THIS FIGURE COMMENTED OUT demonstrates, in this regime much larger amplitudes and higher frequencies are required to twin systems, despite the fact that their potential parameters are proportionately \emph{more} similar.  
% %  \begin{figure}
% \begin{center}
% \includegraphics[width=1\columnwidth]{highU}\end{center}\caption{Twinning fields and response for $U^{(1)}=10t_0$ and $U^{(2)}=(10+0.5)t_0$. Here $\Phi_\tau(t)$ exhibits far greater amplitudes and frequencies as compared to Fig.\ref{fig:lowU}, while the response $J(t)$ is reminiscent of the white light generated when high $U$ Hubbard chains are exposed to transform-limited pulses. Again, atomic units with energy normalised to $t_0$ are used.}
% \label{fig:highU}
% \end{figure}
% }
 \begin{figure}
\begin{center}
\includegraphics[width=0.8\columnwidth]{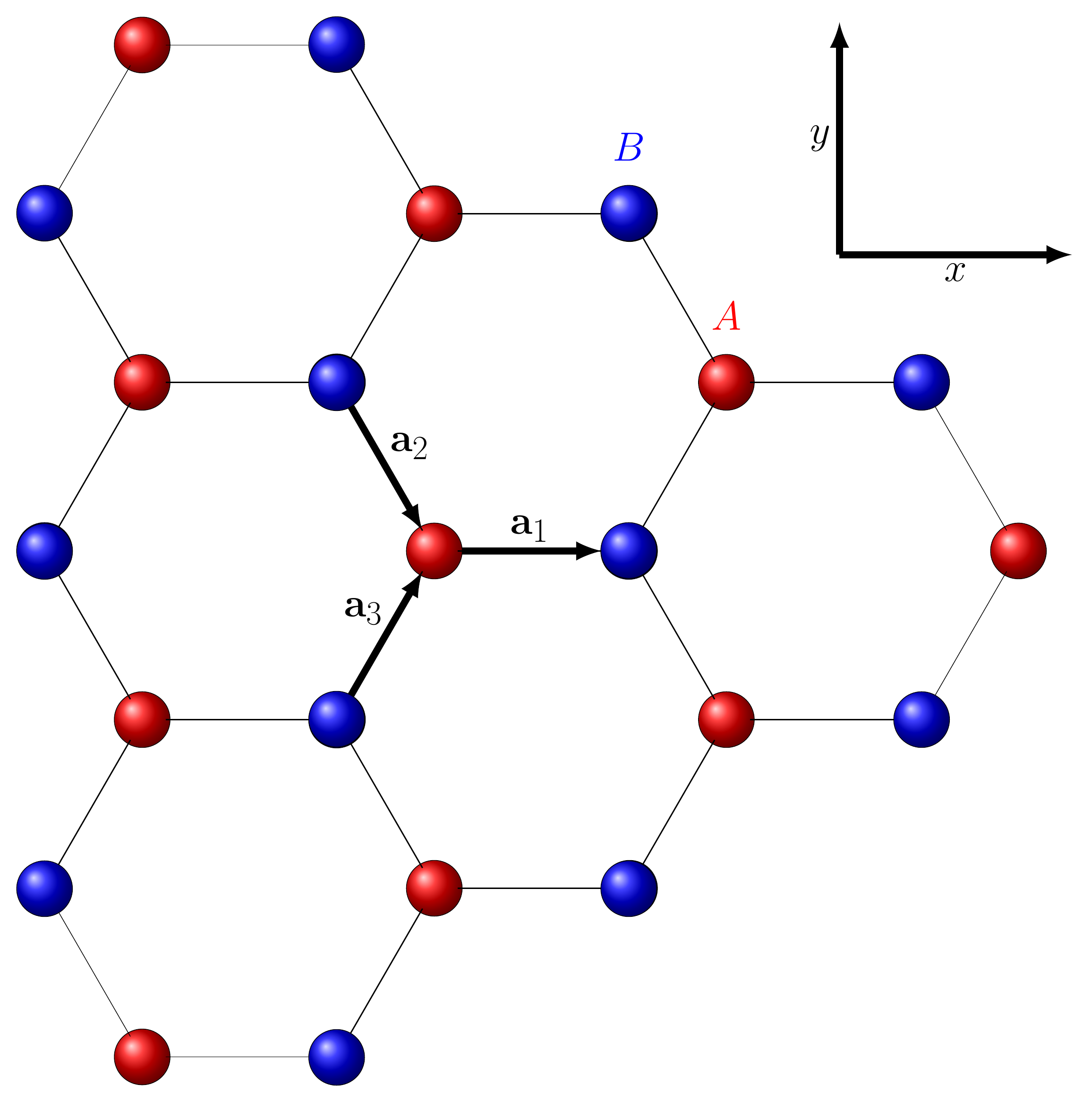}\end{center}\caption{Both graphene and hBN can be modelled with the tight-binding approximation set on a hexagonal bipartite lattice, defined by the positions of the $B$ sublattice atoms relative to those on the $A$ sublattice. These are characterised by the three nearest neighbour vectors $\mathbf{a}_j$, with each being having an angular separation of $120^\circ$, and length 2.5$\AA$ }
\label{fig:lattice}
\end{figure}
\added{
A more concretely physical example is to twin two commonly studied materials - graphene and hexagonal Boron Nitride (hBN). Both of these structures have a bipartite lattice structure, as shown in Fig.\ref{fig:lattice}. Critically, that lattice constant for both systems is almost identical, and can hence be modelled with the identical value $\left|\mathbf{a}^{(gr)}_j\right|=\left|\mathbf{a}^{(hBN)}_j\right|=2.5\AA$ \cite{PhysRevB.76.073103,PhysRevB.102.035441}. This is critical for the existence of a twinning field, as in higher dimensions the ability to twin a pair of systems can only be guaranteed when they share the same lattice structure.  

Both materials are well modelled by the tight-binding approximation \cite{PhysRevB.66.035412}, with an onsite potential of the same form as Eq.~\eqref{eq:onsitepotential}. $t^{(gr)}_0=t^{(hBN)}_0 = 2.7$eV, and for the graphene carbon atoms $U_c=0$, while for hBN $U_B=3.3$eV and $U_N=-1.4$eV for Boron and Nitrogen atoms respectively \cite{,PhysRevB.81.155433,PhysRevB.90.155406}. While it is possible to also include next-to-nearest hopping, the relative strength of this compared to nearest-neighbour hopping is only $\sim 5$\% \cite{RevModPhys.81.109,PhysRevB.76.081406}, and we therefore neglect it for calculational simplicity. Further information on both the precise Hamiltonian describing these systems, and the derived twinning field equations may be found in the supplementary information  \cite{supplement}. 

Simulations are performed using $L=12$ sites with periodic boundaries, and the systems are again prepared via the application of a pumping field. In this case, the polarisation of this initial pump is of great consequence, and in order to generate a physically realisable twinning field, it must be aligned with one of the nearest neighbour vectors. Fig. \ref{fig:twinning2D} shows an example of this, with the initial pump pulse and subsequent twinning field aligned along the $\mathbf{a}_1$ direction. 
 \begin{figure}
\begin{center}
\includegraphics[width=1\columnwidth]{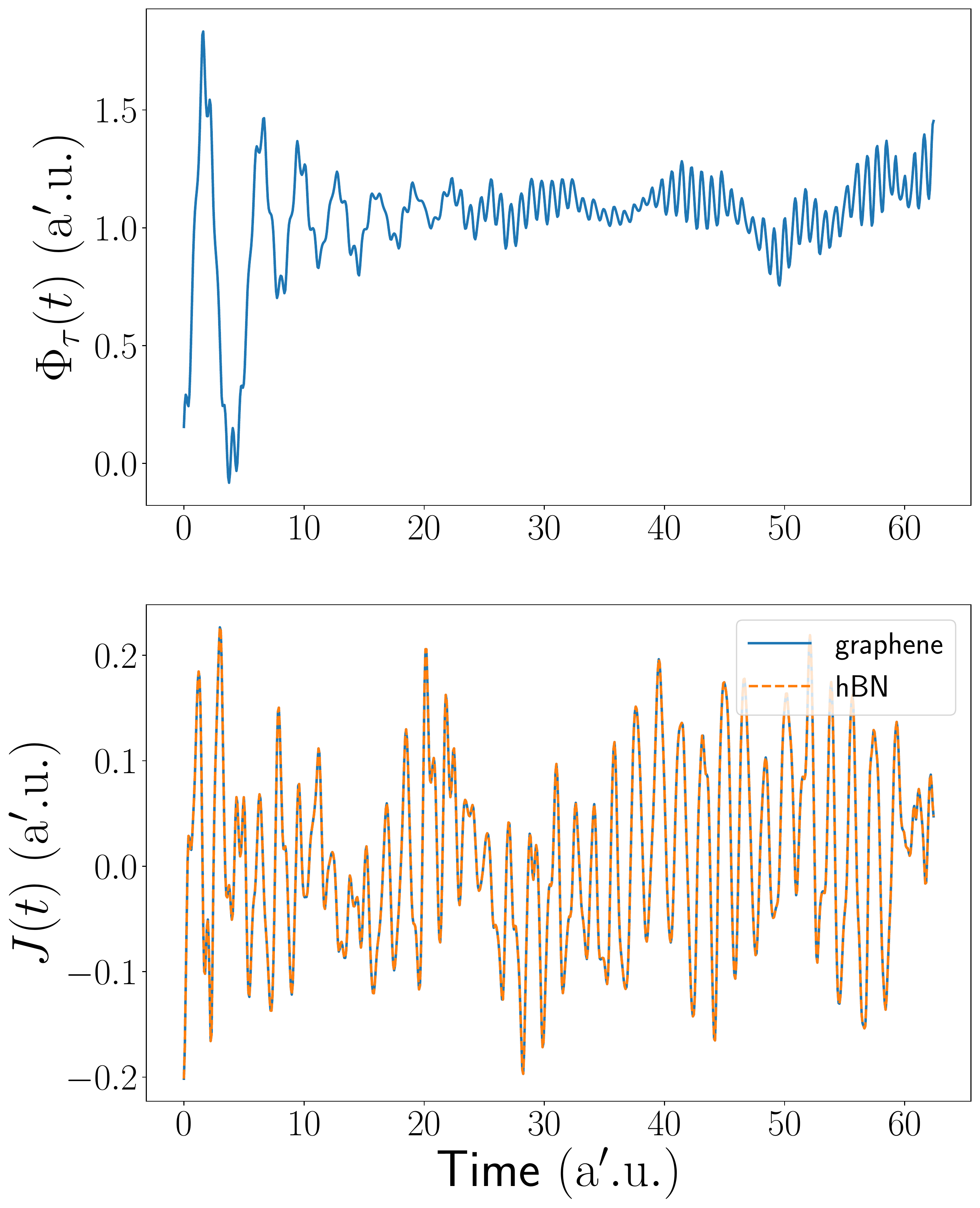}\end{center}\caption{Top panel: Twinning field generated after a pump pulse in the $\mathbf{a}_1$/$x$ direction. The symmetry of the lattice guarantees that the resultant twinning field is \emph{also} purely in the $x$ direction. Bottom panel: the overall current $J(t)$ in the $x$ direction due to application of the twinning field. As expected, this is identical in both systems.  }
\label{fig:twinning2D}
\end{figure}

}
\paragraph*{Discussion:-} We have introduced a non-linear optical phenomenon where \added{some pairs} of quantum systems have a twinning field which generates an identical response in each individual system. \added{In 1D the necessary conditions for a twinning field to exist are rather general, but in higher dimensions the two systems must possess a high degree of similarity in their lattice structures for the existence of a twinning field to be guaranteed.} Conditions for the uniqueness of this field were derived, and numerical calculations provided examples of twinning fields in a Fermi-Hubbard system.  

It is instructive to compare optical indistinguishability with the anti-haecceitism \cite{Saunders2013, sep-haecceitism} of quantum particles. The latter is responsible for both the Fermi-Dirac and Bose-Einstein distributions  \cite{dirac1981the,ZIFF1977169} (as well as the resolution of Gibbs' paradox \cite{Jaynes1965,Jaynes1992}), and is an intrinsic and immutable property of said particles. For this reason it has been commonly assumed that systems governed by distinct Hamiltonians will be distinguishable from each other. Indeed, the effectiveness of spectroscopy is predicated on the notion that a material can be uniquely identified from its spectral response \cite{workman1998applied}. The existence of twinning fields gives the lie to this assumption however, demonstrating that indistinguishability can arise as an emergent property under driving.

It may be tempting to think of twinning fields purely as an act of deception, where an unscrupulous salesman could use the technique to pass off a cheap and nasty material as something more costly. In fact, the analysis presented here demonstrates that it is a trick requiring highly specific conditions to be repeated, and such a fraud can be defeated by an arbitrary modification to the example driving field. Indeed, even if the twinning field is non-unique, the field up to the point that \added{Lipschitz continuity (see \cite{supplement})} is violated \emph{will} be unique. Any field that is distinct from this initial trajectory will therefore be \emph{guaranteed to produce a response distinguishing the two systems}. This has the important consequence of ensuring that techniques designed to discriminate between similar systems are well founded \cite{oddrabitz,li2005optimal, 2010.13859}. 

Of course, the existence of twinning fields forces one to consider both their feasibility and wider utility. \added{While the twinning fields calculated here appear to be rather broadband pulses,} the rapid improvement in both intensity and bandwidth of laboratory laser sources \cite{elu2020seven,lesko2020six} -- combined with the fact that similarly tailored tracking control fields can be well approximated by a few distinct frequencies \cite{tracking2} -- suggests twinning fields may be experimentally realisable with current technology. \added{In fact, deep learning networks have recently been employed to experimentally determine the driving field required to generate a desired response in a material, in a manner that is robust to
noise \cite{Lohani:19}. It is likely that such techniques could be similarly applied for the practical calculation of twinning fields.}  

One potential application of these fields is to characterise the effect of interactions between systems. Applying the twinning field calculated for a non-interacting pair, it would be possible to identify the additional current generated by each system due to its interaction with the other. Twinning fields may also provide a method for creating alternative realisations of metamaterials \cite{RevModPhys.86.1093} when a specific response is required. Given a field and a metamaterial's nearest-neighbour expectation, Eq.~\eqref{eq:twinningfield} can be used to calculate the properties of a different material producing the same effect. Consequently the search for cheaper alternative components in quantum technologies may be aided by twinning fields.

% \deleted{
% Twinning fields can also be used to characterise the effect of interactions between systems. When the systems $\ket{\psi_1}$ and $\ket{\psi_2}$ interact, the individual evolution of each system will be non-unitary. The non-unitary part of the evolution would then generate an additional `coupling current' $\bar{J}^{(k)}$ in each system, such that $J^{(k)}_{\rm tot}(t)=J^{(k)}(t)+\bar{J}^{(k)}(t)$. The operational form of $\bar{J}^{(k)}$ will depend on the form of the inter-system interaction, and in general cannot be expressed as a simple closed form expression, as this would itself require knowledge of the precise master equation describing each system. The unitary nature of the full coupled system evolution does however guarantee that $\bar{J}^{(1)}(t)=-\bar{J}^{(2)}(t)$. By applying the twinning field calculated for the two systems in the non-interacting case, we find that the coupling current is directly observable, as it can be expressed as the difference between each system's total current:}

% \begin{equation}
% \frac{1}{2}\left(J^{(1)}_{\rm tot}(t)-J^{(2)}_{\rm tot}(t)\right)= \bar{J}^{(1)}(t). 
% \end{equation}
%  Once one has obtained the coupling current, a number of methods exist to reconstruct its operator and hence the master equation which generated it \cite{Bairey_2020,PhysRevA.101.062305}. 

Naturally, there remain a number of unanswered questions. For instance, a given twinning field relates a pair of systems, but is this pair unique? Put differently, are there triplets, or $n$-tuplets of systems which exhibit optical indistinguishability? In the case of a non-unique twinning field, what are the physical consequences of choosing one solution over an another? Such questions may merit further investigation, as understanding these secondary properties provides both challenges and opportunities to illuminate the principles upon which driven systems operate.  

\emph{Acknowledgments:}
The authors are grateful to the anonymous referee for comments which have strengthened the paper. The authors would also like to thank Alicia Magann their helpful comments on the manuscript. This work has been generously supported by Army Research Office (ARO) (grant W911NF-19-1-0377; program manager Dr.~James Joseph). The views and conclusions contained in this document are those of the authors and should not be interpreted as representing the official policies, either expressed or implied, of ARO or the U.S. Government. The U.S. Government is authorized to reproduce and distribute reprints for Government purposes notwithstanding any copyright notation herein. This work has also been partially supported by by the W. M. Keck Foundation. 
% \bibliography{refs}
\providecommand{\noopsort}[1]{}\providecommand{\singleletter}[1]{#1}%

\maketitle

%%%%%%% Merge with supplemental materials %%%%%%%%%%
\pagebreak
\widetext
\begin{center}
\textbf{\large Supplemental Material:Optical Indistinguishability via Twinning Fields}
\end{center}
%%%%%%%%%% Merge with supplemental materials %%%%%%%%%%
%%%%%%%%%% Prefix a "S" to all equations, figures, tables and reset the counter %%%%%%%%%%
\setcounter{equation}{0}
\setcounter{figure}{0}
\setcounter{table}{0}
\setcounter{page}{1}
\makeatletter
\renewcommand{\theequation}{S\arabic{equation}}
\renewcommand{\thefigure}{S\arabic{figure}}
\renewcommand{\bibnumfmt}[1]{[S#1]}
\renewcommand{\citenumfont}[1]{S#1}
\renewcommand{\r}{\mathbf{r}}
\newcommand{\rr}{\mathbf{r}^\prime}

\section{Twinning Field Derivation}

\subsection{General Hamiltonian}
Here we provide a full derivation for the twinning field, beginning with the continuum, non-relativistic Hamiltonian for a driven electronic system:
\begin{align}
	\hat{\mathcal{H}}=   \sum_\sigma\int \frac{{\rm d^3}x}{2}\hat{\psi}_\sigma^{\dagger}(\r) \left[-\nabla^2 +V(\r)+\r \cdot \mathbf{E}(\r,t) \right] \hat{\psi}_\sigma(\r) 
		+  \sum_{\sigma \sigma^\prime} \int \frac{{\rm d}^3x {\rm d}^3x'}{2} \hat{\psi}_{\sigma^\prime}^{\dagger}(\r')\hat{\psi}_{\sigma}^{\dagger}(\r) U_{\sigma \sigma^\prime}(\r,\r^\prime) \hat{\psi}_{\sigma}(\r) \hat{\psi}_{\sigma^\prime}(\r')
		\label{eq:GeneralHamiltonian}
\end{align}
Here the  $\hat{\psi}_{\sigma}(\r)$ are standard fermionic field operators satisfying  $\left\{\hat{\psi}_{\sigma^\prime}^{\dagger}(\r'),\hat{\psi}_{\sigma}(\r)\right\}=\delta_{\sigma \sigma^\prime} \delta(\r-\r^\prime)$. $V(\r)$ is an external potential,  while $U_{\sigma \sigma^\prime}(\r,\r^\prime)$ is the two-body interaction potential, and the system is coupled to the driving electric field via the dipole approximation $\r \cdot \mathbf{E}(\r,t)$. Note that in principle it is possible to include higher order interactions without materially affecting the derivation, but in the interests of clarity and concision, we do not include these terms. For similar reasons, we specify that the driving field be homogeneous, i.e. $\mathbf{E}(\r,t)\equiv \mathbf{E}(t)$.

In order to establish a direct relationship between a given expectation and the driving field, the system must be discretised in time or space. Performing a spatial discretisation, we obtain the general form:
\begin{equation}
    \hat{H}=\sum_{\r\rr, \sigma} t(\r,\r^\prime) \hat{c}^\dagger_{\r \sigma} \hat{c}_{\rr\sigma}+\sum_{\r \rr, \sigma \sigma^\prime} U_{\sigma \sigma^\prime}(\r,\r^\prime)\hat{n}_{\r \sigma} \hat{n}_{\rr \sigma^\prime} +\sum_{\r,\sigma} \hat{n}_{\r \sigma}\r\cdot\mathbf{E}(t).
    \end{equation}
In this case one body terms have been collected into (the symmetric function) $t(\r,\rr)$, and the field operators have been replaced by the lattice annihilation operators $\{\hat{c}^\dagger_{\r\sigma},\hat{c}_{\r^{\prime}\sigma^\prime}\}=\delta_{\sigma \sigma^\prime}\delta_{\r \r^{\prime}}$. Finally we have defined the number operator $\hat{n}_{\r \sigma}=\hat{c}^\dagger_{\r\sigma}\hat{c}_{\r \sigma}$. To proceed further we perform the following time dependent unitary transformation of the Hamiltonian: 
\begin{align}
\hat{H} &\to \hat{R}^\dagger(t)\hat{H}\hat{R}(t)-i \hat{R}^\dagger(t)\frac{\partial \hat{R}(t)}{\partial t}\\
    \hat{R}(t)&={\rm e}^{i\r\cdot\mathbf{A}(t)},\ \ \mathbf{A}(t)=\int^t_{-\infty} {\rm d}t^\prime\  \mathbf{E}(t),
\end{align}
from which we obtain:
\begin{equation}
    \hat{H}=\sum_{\r\rr, \sigma}\left( {\rm e}^{-i(\r-\rr)\cdot\mathbf{A}(t)} t(\r,\r^\prime) \hat{c}^\dagger_{\r \sigma} \hat{c}_{\rr\sigma}+{\rm h.c.} \right) +\sum_{\r \rr, \sigma \sigma^\prime}  U_{\sigma \sigma^\prime}(\r,\r^\prime)\hat{n}_{\r \sigma} \hat{n}_{\rr \sigma^\prime}. 
    \end{equation}

At this point, it is helpful to specify the form of the one-body term $t(\r,\rr)$. For the sake of simplicity, we take this to be a nearest neighbour interaction, but note it is possible to include additional next-to-nearest neighbour (and even longer range) hopping terms without affecting our ability to derive a closed form for the twinning field. In the systems we consider here, such terms will be small relative to the nearest neighbour hopping, and their presence would further clutter equations already suffering from index overburdening. 

If each site has $M$ neighbours, described by the relative vectors $\mathbf{a}_j$, then we express the hopping as:
\begin{equation}
    t(\r,\rr)=-t_0 \sum_{j=1}^{j=M}  \delta_{(\r-\rr)\mathbf{a}_j},
\end{equation}
which yields the Hamiltonian
\begin{equation}
    \hat{H}=-t_0\sum_{\r, \sigma}\sum_j\left( {\rm e}^{-i\Phi_j(t)} \hat{c}^\dagger_{\r \sigma} \hat{c}_{(\r+\mathbf{a}_j)\sigma}+{\rm h.c.} \right) + \hat{U},
    \end{equation}
where we have abbreviated $\sum_{\r \rr, \sigma \sigma^\prime} U_{\sigma \sigma^\prime}(\r,\r^\prime)\hat{n}_{\r \sigma} \hat{n}_{\rr \sigma^\prime}\equiv\hat{U}$ and defined the Peierls phase $\Phi_j(t)=\mathbf{a}_j \cdot \mathbf{A}(t)$.

\subsection{General Twinning Field}
Having obtained the appropriate form for the Hamiltonian, we now consider how a twinning field might be derived from it. Let us say we have two systems, described by $\hat{H}^{(1)}$ and $\hat{H}^{(2)}$ each driven by the twinning vector potential $\mathbf{A}_O(t)$, which is designed such that the expectation of some operator $\hat{O}$ in the two systems satisfies  $O^{(1)}(t)=O^{(2)}(t)$. The Hamiltonian for each system will then be given by:
\begin{equation}
  \hat{H}^{(k)}=-t_0^{(k)}\sum_{\r^{(k)}, \sigma}\sum_j\left( {\rm e}^{-i\Phi^{(k)}_j(t)} \hat{c}^{\dagger(k)}_{\r^{(k)} \sigma} \hat{c}^{(k)}_{(\r^{(k)}+\mathbf{a}^{(k)}_j)\sigma}+{\rm h.c.} \right) + \hat{U}^{(k)},   
\end{equation}
where $\Phi^{(k)}_{Oj}=\mathbf{a}^{(k)}_j \cdot \mathbf{A}_O(t)$.
In order to twin $\hat{O}$, it is necessary to match the \emph{derivative} of that expectation if the operator itself has no explicit $\mathbf{A}_O(t)$ dependence. Given $\frac{{\rm d} O^{(k)}}{{\rm d}t}=-i\bra{\psi_k}\left[\hat{H}^{(k)}(t),\hat{O}(t)\right] \ket{\psi_k}$, for an expectation to be twinned we require
\begin{equation}
    \bra{\psi_1}\left[\hat{H}^{(1)}(t),\hat{O}(t)\right] \ket{\psi_1}=\bra{\psi_2}\left[\hat{H}^{(2)}(t),\hat{O}(t)\right] \ket{\psi_2}.
\end{equation}
The conditions under which this equality can be satisfied are best analysed by breaking up the commutator expectations with the following expectations:
\begin{align}
    A_j(\psi_k){\rm e}^{i\theta_j(\psi_k)}&=-t^{(k)}_0\sum_{\r^{(k)},\sigma} \braket{\psi_k| \left[\hat{c}^{\dagger(k)}_{\r^{(k)} \sigma} \hat{c}^{(k)}_{(\r^{(k)}+\mathbf{a}^{(k)}_j)\sigma},\hat{O} \right]|\psi_k}, \\ 
    B^{(k)}(\psi_k)&= \braket{\psi_k| \left[\hat{U}^{(k)}, \hat{O} \right]|\psi_k}.
\end{align}
With some rearrangements, we are able to collect all terms dependent on the driving field, to find a general expression for the twinning field:
\begin{equation}
\label{eq:twinning_general_case}
    \sum_j \left[A_j(\psi_1)\sin\left(\Phi^{(1)}_{Oj}(t)-\theta_j(\psi_1)\right)-A_j(\psi_2)\sin\left(\Phi^{(2)}_{Oj}(t)-\theta_j(\psi_2)\right)\right]= B^{(2)}(\psi_2)- B^{(1)}(\psi_1).
\end{equation}
Clearly such an equation only implicitly defines the twinning field and as such there is no guarantee that a field exists satisfying Eq.~\eqref{eq:twinning_general_case}. If however this equation can be reduced to a closed-form expression for the field, then it is possible to assess when a solution exists. 

\subsection{Twinning Field For Current}
While the existence of a twinning field for a general observable depends on if Eq.~\eqref{eq:twinning_general_case} possesses a solution, there is a particular observable for which twinning fields for a large class of materials can be demonstrated. Specifically, it is the electronic current $\hat{\mathbf{J}}$, which directly determines the system's optical response. This observable is derived directly from the continuity for the electron density $\hat{\rho}=\sum_{\r,\sigma} \hat{n}_{\r \sigma}$. We obtain an expression for this operator by evaluating:
\begin{equation}
    \frac{{\rm d} \hat{n}_{\r \sigma}}{{\rm d}t}=-i\left[\hat{H},\hat{n}_{\r \sigma}\right]=-it_0\left(\sum_{\rr,j,\sigma^\prime}{\rm e}^{-i\Phi_j(t)}\left[ \hat{c}^\dagger_{\rr \sigma^\prime} \hat{c}_{(\rr+\mathbf{a}_j)\sigma^\prime}, \hat{n}_{\r \sigma}\right]-{\rm h.c.}\right)
\end{equation}
Critically, we are able to obtain the second equality due to the fact that the electron-electron interactions captured by $\hat{U}$ depend only on number operators (even if higher order terms are included), and such terms will commute with $\hat{n}_{\r \sigma}$. Evaluating the commutator in the final equality, we obtain:
\begin{equation}
 \left[\hat{c}^\dagger_{\rr \sigma^\prime} \hat{c}_{(\rr+\mathbf{a}_j)\sigma^\prime},\hat{n}_{\r\sigma}\right]=\delta_{\sigma\sigma^\prime}\left(\delta_{\r(\rr+\mathbf{a}_j)}\hat{c}^\dagger_{\rr \sigma^\prime} \hat{c}_{\r\sigma} -\delta_{\r\rr}\hat{c}^\dagger_{\r \sigma} \hat{c}_{(\rr+\mathbf{a}_j)\sigma^\prime}\right).  
\end{equation}
Defining the operator using ($\left|\mathbf{a}_j\right|=a_j$)
\begin{equation}
\hat{J}_{\r j}=-i a_j t_0\sum_{\sigma} {\rm e}^{-i\Phi_j(t)} \hat{c}^\dagger_{\r \sigma} \hat{c}_{(\rr+\mathbf{a}_j)\sigma} -{\rm h.c.}   
\end{equation}
using this operator, we find that the continuity equation can be expressed as
\begin{equation}
 \frac{{\rm d} \hat{\rho}}{{\rm d}t}=\sum_{j} \frac{1}{a_j} \sum_\r \left(\hat{J}_{\r j}-\hat{J}_{(\r-\mathbf{a}_j) j}\right).    
\end{equation}
The right hand side of this equation can immediately be recognised as a discretised divergence, and hence we can identify the current in the $j$ direction as $\hat{J}_j=\sum_\r \hat{J}_{\r j}$.

In the case of this observable the expression for the twinning field ($\Phi^{(k)}_{\tau j}(t)=\mathbf{a}^{(k)}_j\cdot\mathbf{A}_\tau (t)$) is significantly simplified. By first expressing the nearest neighbour hopping expectation in a polar form
\begin{equation}
   \sum_{\r^{(k)},\sigma} \braket{\psi_k|\hat{c}^{\dagger(k)}_{\r^{(k)} \sigma}\hat{c}^{(k)}_{(\r^{(k)}+\mathbf{a}^{(k)}_j)\sigma}|\psi_k}=R_j(\psi_k){\rm e}^{i\theta_j(\psi_k)}
\end{equation}
The current expectation in the $j$ direction may be given as
\begin{equation}
    J^{(k)}_j(t)=\braket{\psi_k|\hat{J}_j|\psi_k}=2a^{(k)}_j t^{(k)}_0 R_j(\psi_k) \sin(\Phi^{(k)}_{\tau j}(t)-\theta_j(\psi_k)).
\end{equation}
Equating $\hat{J}_j^{(1)}=\hat{J}_j^{(2)}$, we obtain
\begin{equation}
\label{eq:almost!}
 a^{(1)}_j t^{(1)}_0 R_j(\psi_1) \sin(\Phi^{(1)}_{\tau j}(t)-\theta_j(\psi_1))-a^{(2)}_j t^{(2)}_0 R_j(\psi_2) \sin(\Phi^{(2)}_{\tau j}(t)-\theta_j(\psi_2)) =0.   
\end{equation}
Comparing to Eq.~\eqref{eq:twinning_general_case}, we find two major simplifications that bring us close to a closed form equation for the field. First, the twinning field does not explicitly depend on any expectation of the potential $\hat{U}^{(k)}$ term (although the evolution of the state will depend on this potential). Secondly, the specific form of the current operator means that each component of the current in the two systems can be equated separately, meaning one has an equation for the twinning field in each of the $j$ nearest neighbour directions, rather than a single equation summing over all components.

Unfortunately, it is still not possible to obtain a closed form equation for the twinning field from Eq.~\eqref{eq:almost!}. To amend this we make the further assumption that both systems have the same lattice structure, $\r^{(1)}=\r^{(2)}$ and $\mathbf{a}^{(1)}_j=\mathbf{a}^{(2)}_j$. In this case $\Phi^{(1)}_{\tau j}(t)=\Phi^{(2)}_{\tau j}(t)\equiv \Phi_{\tau j}(t)$, and Eq.~\eqref{eq:almost!} can be rearranged to express the twinning field directly in terms of the systems' expectations:
\begin{align}
    \Phi_{\tau j}(t) &=\arctan\left(\xi_j(\psi_1,\psi_2)\right) \label{eq:3Dtwinningfield}\\ 
    \xi_j(\psi_1,\psi_2)&=\frac{\lambda R_j(\psi_1)\sin(\theta_j(\psi_1))-R_j(\psi_2)\sin(\theta_j(\psi_2))}{ \lambda R_j(\psi_1)\cos(\theta_j(\psi_1))-R_j(\psi_2)\cos(\theta_j(\psi_2))},
\end{align}
using $\lambda=\frac{t^{(1)}_0}{t^{(2)}_0}$. Note that both $R_j(\psi_k)$ and $\theta_j(\psi_k)$ will have definite values for every valid state $\ket{\psi_k}$ in each system. If this were not the case then the action of the nearest neighbour hopping operator (and therefore the Hamiltonian) on the system would not be well defined. For this reason, $\xi_j(\psi_1,\psi_2)$ will have a definite real value regardless of the states of the system. Given the $\arctan$ function maps any real to $(-\pi/2, +\pi/2)$, we conclude that $\Phi_{\tau j}$ will always have a definite value.  

To summarise, while an expression for the twinning field can always be obtained, in general it cannot be ascertained as to whether this equation will possess a solution. \emph{If the two systems share an identical lattice structure then it it possible to express the twinning field in a closed form.} Under these conditions, the twinning field may be explicitly expressed as a function of well-defined system expectations, and given that the twinning field is a function whose range and domain extends over the entire real line, it is guaranteed to exist regardless of the value of those system expectations. This leads to the perhaps surprising conclusion that ultimately it is the lattice structure of a material, rather than its electronic interactions, that determine if its optical response can be twinned.

\subsection{Specification to Hexagonal Bipartite Lattice}
Here we outline the notation used for the twinning field for a hexagonal bipartite lattice (shown in figure 3), in order to simulate graphene and hBN. In this case, the bipartite nature of the lattice means that the sum over lattice positions can be indexed directly, where sites belonging to the $A$ and $B$ sublattices are contained in the sets $\mathcal{A}$ and $\mathcal{B}$ respectively. If an onsite potential is used, the Hamiltonian reads:

\begin{equation}
    \hat{H}=-t_0\sum_{<i,j>, \sigma}\left({\rm e}^{-i\Phi_j(t)}\hat{c}_{i\sigma}^\dagger\hat{b}_{j\sigma}+{\rm h.c.}\right) +\sum_{i\in\mathcal{A}}U_A\hat{n}_{i\uparrow}^A \hat{n}_{i\downarrow}^A+\sum_{j\in\mathcal{B}}U_B\hat{n}_{j\uparrow}^B \hat{n}_{j\downarrow}^B
\end{equation}
here $\hat{c}$ operators act on the $A$ sublattice, while $\hat{b}$ acts on the $B$ sublattice. Each sublattice possesses a separate onsite potential $U_A$ and $U_B$, using $\hat{n}_{i\sigma}^{A/B}$ as the number operators acting on the appropriate lattice. Finally, each site has three neighbours related by the vectors:
 \begin{align}
    \mathbf{a}_1 =a\begin{bmatrix}
           1 \\
           0 
         \end{bmatrix},
\ \ \mathbf{a}_2 =\frac{a}{2}\begin{bmatrix}
           1 \\
           -\sqrt{3} 
         \end{bmatrix},
 \ \ \mathbf{a}_3 =\frac{a}{2}\begin{bmatrix}
   1 \\
   \sqrt{3} 
 \end{bmatrix}.
  \end{align}
 A current in each of these directions can be defined in the same manner as the previous section:
 \begin{equation}
   \hat{J}_j = -ia t_0\sum_{i\in \mathcal{A}, \sigma}\left({\rm e}^{-i\Phi_j(t)}\hat{c}_{i\sigma}^\dagger\hat{b}_{j\sigma}-{\rm h.c.}\right).
 \end{equation}
 Then, if we once again define the functionals for each separate system,
 \begin{equation}
   \sum_{i\in \mathcal{A}^{(k)},\sigma} \braket{\psi_k|\hat{c}^{\dagger(k)}_{i\sigma}\hat{b}^{(k)}_{j\sigma}|\psi_k}=R_j(\psi_k){\rm e}^{i\theta_j(\psi_k)},
\end{equation}
it is possible to apply Eq.~\eqref{eq:3Dtwinningfield} directly.

Even in this case (where each component of a twinning field has a closed form), it is important to remark that in general this does not mean the fields generated are physically realisable. To see this, consider the fact that one has three independent equations for the $\Phi_j(t)$, but geometrically there are only two independent components. In Cartesian coordinates, the fields in the three lattice directions are given by:
\begin{equation}
    \Phi_{\tau1}(t)=\Phi_{\tau x}(t), \ \ \Phi_{\tau2}(t)=\frac{1}{2}\left(\Phi_{\tau x}(t)+\sqrt{3}\Phi_{\tau y}(t)\right), \ \ \Phi_{\tau3}(t)=\frac{1}{2}\left(\Phi_{\tau x}(t)-\sqrt{3}\Phi_{\tau y}\right).
\end{equation}
While it is always possible to find a solution for the fields in the lattice directions, there is no guarantee that these will be consistent with the geometric composition of these fields. Fortunately, when the field is polarised along one of the lattice directions (say $\mathbf{a}_1$), the symmetry of the system guarantees that currents in the two remaining directions will be equivalent, i.e. $J_2(t)=J_3(t)$, such that $J_y(t)=0$. Furthermore, local current conservation means that in this case $J_1(t)=J_2(t)+J_3(t)$. Consequently, the total current in the $x$ direction will be $J_x(t)=2J_1(t)$.  This means only the  component of the twinning field in that direction need be calculated, removing the aforementioned geometric constraints.   

\subsection{Uniqueness of Twinning Fields}
While a solution for the twinning field may exist, the  existence of this solution says nothing about the \emph{number} of twinning fields that could exist for a given pair of systems -- i.e. is a twinning field unique? It is possible to deduce an answer to this question by substituting  $\Phi_\tau(t)$  into the Hamiltonian for each system. In this case we obtain an explicitly nonlinear evolution governed by a `twinning Hamiltonian' for the joint system (here using a 1D example):
\begin{align}
\hat{H}^{\rm tot}_\tau(\psi_1,\psi_2)&=\hat{H}_\tau^{(1)}(\psi_1,\psi_2)\otimes \mathbb{1}^{(2)} +\mathbb{1}^{(1)}\otimes \hat{H}_\tau^{(2)}(\psi_1,\psi_2) \label{eq:fulltwinningHamiltonian} \\ 
\hat{H}^{(k)}_\tau\left(\psi_1,\psi_2 \right) & = \sum_{\sigma,j} \left[ P^{(k)}_+\hat{c}_{j\sigma}^{\dagger}\hat{c}_{j+1\sigma} + P^{(k)}_-\hat{c}_{j\sigma}\hat{c}^\dagger_{j+1\sigma} \right] + \hat{U}^{(k)}, \label{eq:twinningHamiltonian} \\
P^{(k)}_{\pm} & = -\frac{t^{(k)}_{0}}{\sqrt{1 + \xi^2(\psi_1,\psi_2)}}\left(1 \pm i \xi(\psi_1,\psi_2)\right) \label{eq:PlusMinus}.
\end{align}

Using this twinning Hamiltonian, we can assess the conditions under which the twinning field is unique. If the evolution generated by the Hamiltonian in Eq.~\eqref{eq:fulltwinningHamiltonian} from a given initial condition has only one solution, it implies there is only one field which satisfies the twinning condition. Applying Lipschitz analysis \footnote{In order to assess the uniqueness of the evolution of the joint state $\ket{\Psi}=\ket{\psi_1}\otimes\ket{\psi_2}$ under $\hat{H}^{\rm tot}_\tau$, we make use of the Picard-Lindel{\"o}f theorem \cite{kentnagle2011}. This states that the solution to a differential equation $y^\prime(t)=f(t,y(t)), y(t_0)=y_0$ is unique if $f(t,y(t))$ is a \emph{Lipschitz continuous} (LC) \cite{geraldfolland2007} function over  all $y$. It therefore follows that the conditions under which $\Phi_\tau (t)$ is unique are equivalent to those for which $\hat{H}^{\rm tot}_\tau\ket{\Psi}$ is LC.}  to this Hamiltonian, it is possible to prove the following: 

\emph{For a finite dimensional system (i.e., when the twinning Hamiltonian $\hat{H}^{\rm tot}_\tau$ is a finite dimensional matrix), if $\ket{\Psi}$ solves Eq.~\eqref{eq:fulltwinningHamiltonian} and satisfies the constraints}
\begin{align}
\left|R(\psi_{1/2})\right|>&\epsilon_1>0, \label{eq:Lipschitz1} \\
\left|{\rm Re}\left(\lambda K(\psi_1)-K(\psi_2)\right)\right|>&\epsilon_2>0, \label{eq:Lipschitz2}
\end{align}
\emph{where $\epsilon_{1/2}$ are any constants, then the twinning field $\Phi_\tau(t)$ is unique.}

The first condition [Eq.~\eqref{eq:Lipschitz1}] can be straightforwardly interpreted as the requirement that there be some kinetic energy in each system, a natural precondition for observing a current. Some sense can be made of the second condition by noting that evolution under the twinning field implies $\frac{{\rm d}J^{(1)}(t)}{{\rm d} t}=\frac{{\rm d}J^{(2)}(t)}{{\rm d} t}$. Evaluating this derivative, we obtain: 
\begin{align}
    \dot{\Phi}_\tau(t) {\rm Re}\left(\lambda K(\psi_1)-K(\psi_2)\right)=  {\rm Re}&\left( \left\langle \psi_1 (t) \left|\left[\hat{U}^{(1)},\hat{K}\right]\right| \psi_1 (t)  \right\rangle\right) 
-{\rm Re}\left( \left\langle \psi_2 (t) \left|\left[\hat{U}^{(2)},\hat{K}\right]\right| \psi_2 (t)  \right\rangle\right).    
\end{align}
Clearly, when Eq.~\eqref{eq:Lipschitz2} is violated, $\dot{\Phi}_\tau$ can take any value while still functioning as a twinning field. This means that in principle, if a twinning field is non-unique, there are an \emph{infinite} number of solutions. These solutions will however be constrained to follow the same trajectory up to the point where one of Eq.~\eqref{eq:Lipschitz1} or Eq.~\eqref{eq:Lipschitz2} is violated, and a proliferation of possible trajectories occurs. This non-uniqueness is therefore qualitatively different to that previously observed in control fields for tracking control \cite{tracking2,Rabitzmultiple},  where a branch point generates a well-defined number of additional solutions.  
% \bibliography{refs}

\providecommand{\noopsort}[1]{}\providecommand{\singleletter}[1]{#1}%

\end{document}